# A Dimensional Approach to Canine Bark Analysis for Assistance Dog Seizure Signaling

Hailin Song, Shelley Brady,
Tomás Ward, and Alan F. Smeaton[

Dublin City University, Glasnevin, Dublin 9, Ireland
alan.smeaton@dcu.ie

**Abstract.** Standard classification of canine vocalisations is severely limited for assistance dogs, where sample data is sparse and variable across dogs and where capture of the full range of bark types is ethically constrained. We reframe this problem as a continuous regression task within a two-dimensional arousal-valence space. Central to our approach is an adjusted Siamese Network trained not on binary similarity, but on the ordinal and numeric distance between input sample pairs. Trained on a public dataset, our model reduces *Turn-around Percentage* by up to 50% on the challenging valence dimension compared to a regression baseline. Qualitative validation on a real-world dataset confirms the learned space is semantically meaningful, establishing a proof-of-concept for analysing canine barking under severe data limitations.

## 1 Introduction

Epilepsy is a common neurological disorder where seizures are often drug-resistant, impacting millions of people worldwide [1-4]. Assistance dogs have emerged as a promising aid for people who have seizures, capable of detecting their pre-ictal Volatile Organic Compounds (VOCs) through canine olfaction [5, 6]. Evidence shows that even untrained dogs can alert to seizures in their owners, while trained dogs can do so with high accuracy and significant warning times [7, 8]. Despite these benefits, the inherent stress of seizure events in their owners raises important animal welfare considerations, as adverse health effects in responding dogs have been reported [9].

A challenge in canine detection and alerting of seizures is the design of the dogs' signaling behaviours, which need to consider animal welfare. While artificial signals like a spinning movement can be automatically detected [10], they may conflict with dogs' innate stress responses namely staring or barking. Adhering to animal-computer interaction principles [11], this work investigates leveraging a natural vocalisation, the attention-seeking bark, as a trained and automatically detectable seizure alert.

While supervised bark classification is well-established, leveraging large public datasets [12-15], these data-intensive methods are severely limited for our application of training a model on a single, known assistance dog. The rigorous data collection and annotation processes required [16] cannot be replicated at scale for an individual animal due to severe practical and ethical constraints. Barking data for one dog is limited to a short collection period, during which it is ethically prohibited to artificially induce high-stress scenarios to elicit a full range of bark types. This real-world context yields scarce and incomplete datasets for each dog, fundamentally shifting our challenge from standard classification to developing a robust detection method that can operate under these severe training data limitations.

To overcome the limitations of categorical classification, this paper reframes canine vocalisation analysis as a regression task, proposing a framework to map barks onto a continuous, two-dimensional arousal-valence space from limited and ordinally-labeled data. Central to our approach is an adjusted Siamese Network. Unlike canonical Siamese architectures that learn a binary similarity metric, our model is optimised for regression using an ordinal and numeric distance objective, allowing it to explicitly model the ordered relationships within the data (e.g., 'High' > 'Medium' > 'Low') as a psychologically-grounded dimensional representation. We validate the utility of our emotional space by projecting real-world canine barks from 7 dogs onto the learned manifold. While not comprehensive, this is a proof-of-concept, demonstrating the framework's potential for identifying signaling behaviours for applications such as seizure alerting.

## 2     Related Work

### 2.1    Categorisation of Canine Barking

The predominant approach to categorising canine vocalisations has been context-driven classification, where discrete labels are derived from controlled situational stimuli (e.g., 'Stranger,' 'Fight,' 'Play') [12, 13, 17]. This presents limitations in granularity and specificity. For instance, distinct contexts like 'Ball' and 'Food' can elicit acoustically similar barks, making the labels ambiguous [17] and creating a need for secondary emotion scales ('Aggressiveness', 'Playfulness') [13].

These limitations have motivated a shift towards a dimensional model of emotion, drawing from foundational work in animal psychology [18, 19, 16]. This approach characterises affective states by their position within a continuous two-dimensional arousal-valence space (Fig. 1), defined by arousal (intensity) and valence (hedonic value). While this concept has been proposed previously, its application in machine learning work has not extended to direct, continuous prediction.

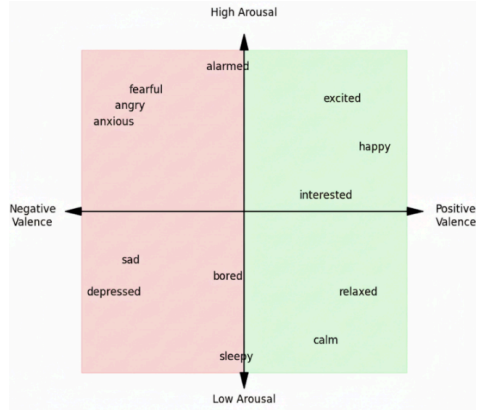

**Fig. 1.** Two dimensional arousal-valence space

### 2.2 Feature Extraction and Classification Models in Canine Bark Analysis

A conventional machine learning pipeline for canine bark analysis comprises data acquisition, feature extraction, model training, and finally classification. A significant body of research has focused on applying this pipeline to classify barks based on context or demographics such as dog breed and age [12, 15, 16, 17, 20, 21].

For feature extraction in bioacoustics, representations that model perceptual hearing, such as the Mel spectrogram and Mel-Frequency Cepstral Coefficients (MFCCs), have become the *de facto* standard, providing a compact representation of an audio signal's spectral properties [21]. These features (see Fig. 2) are well-suited as input for Convolutional Neural Networks (CNNs), which are widely used for their ability to learn hierarchical features by treating spectrograms as images [22]. As the primary objective of our work is to reformulate the task from discrete classification to continuous dimensional regression, we employ this standard and well-validated approach, using spectrogram features as input to a foundational CNN structure.

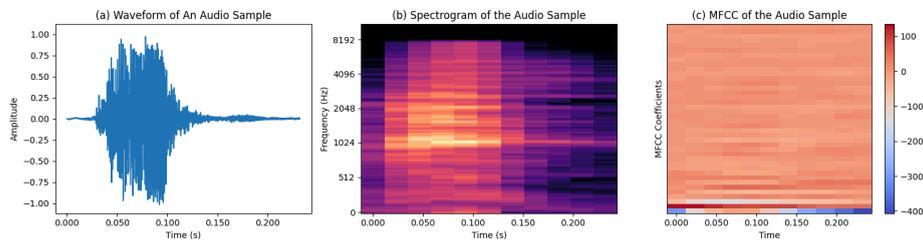

**Fig. 2.** Visualisation of the Waveform, Spectrogram and MFCC of a canine bark sample.

### 2.3 Siamese Neural Networks for Feature Space

The Siamese Neural Network is a class of architectures designed for metric learning, where the primary objective is to learn a meaningful, low-dimensional embedding space rather than perform direct classification [23]. By processing input pairs through identical, weight-sharing subnetworks (see Fig. 3), it learns a generalisable similarity function that pulls similar samples closer together while pushing dissimilar ones apart. This is highly effective for tasks with limited data per class such as one-shot image recognition [24] where the aim is to find a discriminative, clustering-friendly embedding space where proximity reflects semantic similarity [25].

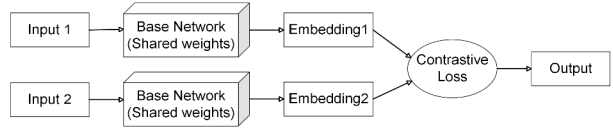

**Fig. 3.** Regular Siamese network architecture.

The canonical Siamese architecture is optimised for a binary similarity objective (i.e., 'similar' vs. 'dissimilar') which is insufficient for our task. Our framework relies on ordinal annotations for arousal and valence (e.g. 'High', 'Medium', 'Low'), where the directional relationship between labels is crucial. To address this, we introduce an adjusted Siamese Network to learn a continuous scalar value that preserves ordinal constraints. We validate its efficacy in learning a continuous dimensional representation from sparsely-labeled, ordinal data which is variable across dogs.

## 3     Methodology and Materials

### 3.1    Two-Dimensional Emotional Space

Inspired by the circumplex model of affect (Fig. 1), our methodology projects canine vocalisations onto a continuous two-dimensional emotional space defined by the orthogonal axes of arousal and valence. Our pipeline (Fig. 4) begins with feature extraction, where each bark audio segment is converted into a spectrogram-based feature representation. We then employ two independently trained models to predict a continuous scalar value for each coordinate. The outputs of these models are then combined to form a (valence, arousal) coordinate which is projected onto the emotional space. This allows for a quantitative evaluation of the bark's relative emotional category based on its position within the quadrants of the output.

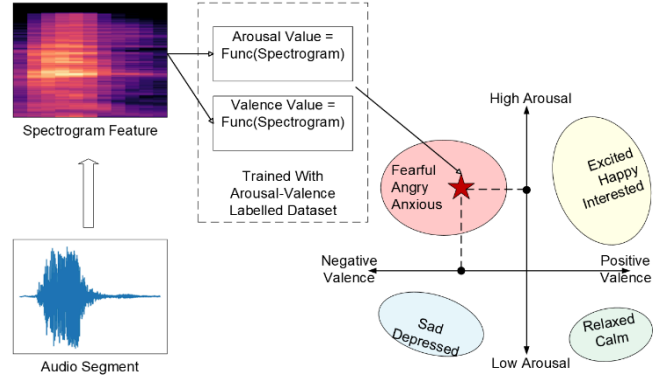

**Fig. 4.** Projection of audio to emotional space

### 3.2    Data Segmentation and Feature Extraction

To account for the heterogeneous nature of continuous recordings, we employ an energy-based method, extracting non-silent segments using a threshold of 20 dB below the peak amplitude. The resulting variable-length segments are subsequently normalised to a fixed size of 5120 samples, a length determined by the median segment duration in our dataset. Segments shorter than this length are symmetrically zero-padded, while longer ones are framed using a sliding window with 50% overlap (a stride of 2560 samples). Finally, a Mel spectrogram is computed for each fixed-length segment, serving as the input feature representation for our models.

### 3.3    Model and Evaluation

As our work reframes the task from discrete classification to continuous regression, we establish a standard CNN baseline (Fig. 5) and for our model which takes a Mel spectrogram as input and is trained to directly regress the continuous numerical value corresponding to an ordinal label. Our model is an adaptation of the Siamese architecture (Fig. 3) and uses the same base CNN as its weight-sharing network but its training objective is fundamentally altered. Instead of learning a binary similarity metric, the network is trained as a regressor to predict the ordered numeric distance between the two input samples. For instance, if 'Input 1' has a label of 'High' (1.0) and 'Input 2' has a label of 'Low' (-1.0), the network's target output is their numeric difference, 1.0 - (-1.0) = 2.0.

To facilitate the regression task, the ordinal labels for both arousal and valence were standardised to a numeric scale: 'High'/'Positive' to 1.0, 'Medium'/'Neutral' to 0.0, and 'Low'/'Negative' to -1.0. To mitigate the risk of data leakage and performance inflation from acoustically similar, overlapping frames, the train-test split was performed at the event level. This ensures that all segments derived from a single, continuous vocal event reside exclusively in either the training or the test set.

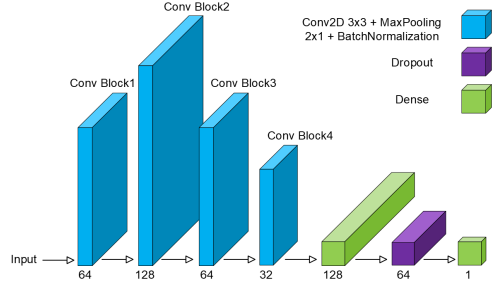

**Fig. 5.** Baseline CNN Architecture.

Given our focus on learning an ordinally-structured space, our evaluation employs two distinct metrics. ***Accuracy*** provides a uniform comparison with traditional classification approaches and the continuous regression outputs from our models are mapped back to their nearest categorical labels ('High'/'Positive', 'Medium'/'Neutral', 'Low'/'Negative') using decision boundaries learned from the distribution of regression values on the training set. Standard accuracy is fundamentally inadequate for ordinal categories, as it treats all misclassifications as equal and fails to penalize severe errors, such as confusing 'High' and 'Low' states, more heavily than minor ones. In learning a meaningful dimensional space, preserving the ordinal relationship is paramount; a model that frequently confuses the extremes of a dimension has failed to learn the data's underlying structure. To provide a more nuanced, task-relevant evaluation, we introduce the ***Turn-around Percentage (TAP)***:

$$\text{TAP} = \frac{N(\text{High} \to \text{Low}) + N(\text{Low} \to \text{High})}{N(\text{Total High}) + N(\text{Total Low})} \quad (1)$$

As defined in Eq. (1), TAP is specifically designed to quantify the most critical type of ordinal error by measuring the proportion of instances misclassified from one extreme of the scale to the other, relative to the total number of extreme instances.

All data processing, model training, evaluation, and visualisation were conducted within the Google Colab cloud environment. The implementation relied on standard Python libraries, primarily using TensorFlow and Keras for model development and Librosa for audio feature extraction.

### 3.4  Datasets

To facilitate a comprehensive evaluation, we used two datasets: a publicly available, dimensionally-annotated corpus for model training and a bespoke, real-world collection for validation.

For the primary task of learning the dimensional emotional space, we used the Emotional Canines dataset [16]. This public corpus is the largest of its kind annotated for dimensional analysis, containing 1,400 bark sequences from Siberian Husky and Shiba Inu breeds. The dataset, recorded at 22.05kHz, is pre-partitioned to prevent subject overlap, with a training set of 600 sequences and a test set of 100 sequences per breed. Each vocalisation is annotated with continuous labels for arousal and valence, and the sequences vary in duration from 0.06 to 30.02 seconds (mean:1.50s, median: 0.96s). Its dimensional annotation scheme makes it uniquely suited for the regression-based task central to our study.

For validation and to assess real-world applicability, we collected a bespoke corpus containing over 454 hours of audio from seven pet dogs. Data was captured non-intrusively via collar-attached miniature voice-activated recorders (e.g., STTWUNAKE, DizerLin) as single-channel (mono) audio at 16 kHz with 16-bit depth. A key characteristic of this "in-the-wild" dataset, summarised in Table 2, is the inherent sparsity of vocalisation events in a natural domestic environment. Across all subjects, the cumulative duration of identified bark events was 2183 seconds, accounting for only 0.133% of total recording time. This extreme class imbalance underscores the "in-the-wild" nature of the data and highlights the challenge of robustly detecting rare yet significant signaling behaviors, which is a central motivation for our proposed framework.

**Table 1.** Descriptive Statistics of the Real-World Canine Bark Dataset.

| Dog Name | Recording Length (hh:mm:ss) | Bark Length (s) | Bark/Recording Ratio (%) |
|---|---|---|---|
| Bear | 16:42:33 | 2 | 0.025 |
| Bertie | 8:10:50 | 11 | 0.336 |
| Connie | 37:23:51 | 6 | 0.108 |
| Lexi | 248:36:46 | 716 | 0.080 |
| Mary | 29:36:10 | 853 | 0.800 |
| Rosie | 65:18:55 | 311 | 0.132 |
| Woody | 49:05:05 | 44 | 0.025 |
| Total | 454:54:10 | 2183 | 0.133 |

## 4    Experimental Results

### 4.1    Accuracy and Turn-Around Percentage

All models were trained and tested on the Emotional Canines dataset and we established multiple baselines for a comprehensive comparison. We refer to the breed-specific classification accuracies originally reported in [16] and to provide a more direct comparison with our regression-based methods, we also trained a standard CNN classification model using the architecture from Fig. 5. Both our Adjusted Siamese Network and a baseline CNN Regressor were trained on the combined two-breed dataset to learn a generalised emotional space. The performance of all models on the arousal and valence prediction tasks is summarised in Table 2.

**Table 2.** Performance comparison on the Emotional Canines test set. The results from [16] and our CNN (Classification) baseline are reported separately for the Husky and Shiba Inu breeds. The regression models were trained on the combined dataset.

| Model | Arousal | | Valence | |
|---|---|---|---|---|
| | Accuracy(%) | TAP(%) | Accuracy(%) | TAP(%) |

| Reported by [16] | 42 / 46 | - | 34 / 45 | - |
| --- | --- | --- | --- | --- |
| CNN(Classification) | **60.45** / 43.35 | 15.37 / 22.87 | **56.34** / 31.91 | 56.34 / 31.91 |
| CNN(Regressor) | 47.85 | 14.41 | 36.70 | 30.56 |
| Adjusted Siamese | 55.93 | **10.83** | 48.07 | **15.91** |

### 4.2   **Visualisation of Predicted Dimensional Distributions**

To assess how well each model learned the ordinal structure of the emotional space, we show the distribution of their continuous regression outputs on both the training and test sets. Figs. 6-7 illustrate the predicted value distributions for the arousal and valence dimensions for the baseline CNN and our adjusted Siamese Network.

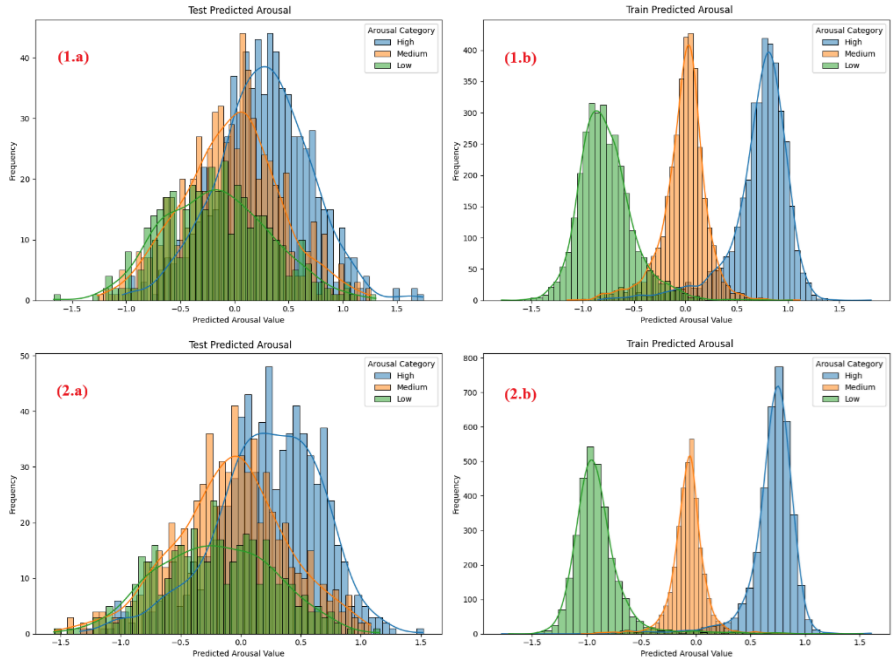

**Fig. 6.** Distributions of predicted arousal values for the baseline CNN (1) and adjusted Siamese Network (2) on the test (a) and train (b) sets.

As shown in Fig. 6, both models learn to effectively separate the three arousal categories on the training set, mapping them to distinct regions along the axis. On the test set, this separation is less pronounced for both models, a finding consistent with expected generalisation performance. The adjusted Siamese Network (Fig. 6(2.a)) maintains a clearer ordinal separation compared to the baseline CNN (Fig. 6(1.a)).

The distributions for the valence dimension, shown in Fig. 7, reveal a more pronounced difference between models. While the Adjusted Siamese Network (Fig. 7(2.a)) again shows a clear, albeit overlapping, separation of the three categories on the test set, the baseline CNN (Fig. 7(1.a)) struggles. The distributions for the valence categories for the baseline model largely collapse into a single mode on the test set, indicating a failure to generalise the ordinal structure for valence. This supports the quantitative results, suggesting that the training objective of the adjusted Siamese Network makes it more capable of learning a robust, ordinally-structured embedding, particularly for the more challenging valence dimension.

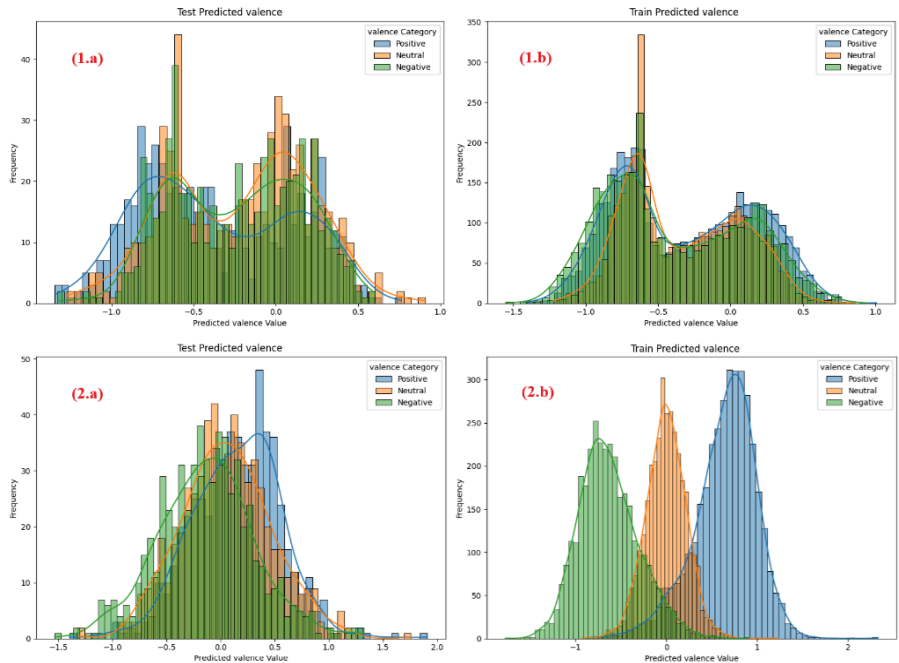

**Fig. 7.** Distribution of predicted valence values for the baseline CNN (1) and adjusted Siamese Network (2) on the test(a) and train(b) sets.

### 4.3  Real-World Canine Bark Dataset

To assess the generalisability of our model, we conducted a qualitative validation using the Real-world Canine Bark Dataset, a corpus characterised by the extreme sparsity of vocalisation events (0.133% of >454 hours)[1]. This section presents a case study analysis, projecting specific, owner-identified bark events onto the emotional space learned from the curated dataset to evaluate the semantic validity of their resulting coordinates.

Following the protocol detailed in Section 3.2, the 2183-seconds of canine barks was segmented and processed to extract spectrogram features for all seven dogs. We then used our trained adjusted Siamese Network to predict the arousal and valence coordinates for each bark. The resulting projections, which map each real-world vocalisation to a point in the learned emotional space, are shown in Fig. 8.

---

[1] This dataset is available at https://doi.org/10.6084/m9.figshare.30137845.v1

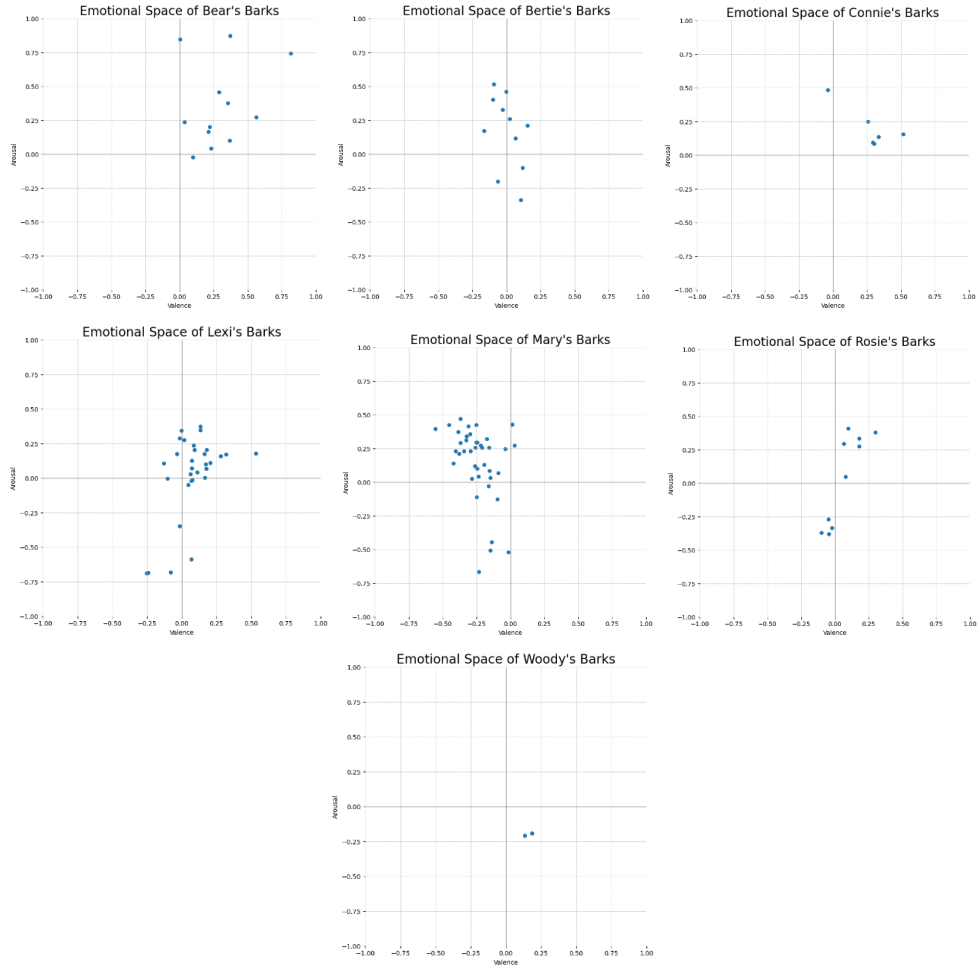

**Fig. 8.** Projection of barks from seven dogs onto the learned arousal-valence space.

A qualitative analysis of these projections reveals a strong correspondence between the barks' positions in the learned space and the dogs' known behavioral contexts. For instance two distinct vocalisations from 'Woody', identified as soft snores during sleep, are projected as proximate points in the low-arousal, positive-valence quadrant (calm/relaxed), aligning with their passive context. In contrast, the majority of barks from 'Connie', a puppy known for frequent attention-seeking barks, are concentrated in the high-arousal, positive-valence quadrant, consistent with an excited or interested state. Conversely, vocalisations from 'Mary', who exhibits a strong tendency to bark at slight noises outside, are predominantly located in the high-arousal, negative-valence quadrant, which corresponds to an anxious or alarm state.

While the Real-world Canine Bark Dataset is inherently sparse and variable across dogs and it lacks precise ground-truth arousal and valence labels, this case-study analysis provides strong qualitative evidence for the validity of our learned emotional space. The consistent alignment between the predicted coordinates and the dogs' known behavioral characteristics suggests that our framework successfully generalises from the curated training data to capture meaningful emotional information for dogs outside the training dataset in a challenging, real-world setting.

## 5   Discussion

Reframing canine vocalisation analysis as dimensional regression proves to be a viable strategy, enabling the creation of a more ordinally coherent feature space without sacrificing traditional accuracy. This is evidenced by the significantly lower Turn-around Percentage (TAP) of our adjusted Siamese Network, which, while achieving comparable accuracy to the baseline, demonstrates superior performance in capturing the intrinsic ordinal structure of the emotional space—the central goal of our work. As detailed in Table 2, this advantage is particularly pronounced in the challenging valence dimension, where the TAP was reduced from 30.56% to 15.91%. This quantitative success is further supported by our qualitative analysis of the private "in-the-wild" dataset, which, despite the extreme sparsity of vocalisations, shows a semantically meaningful alignment between known behavioral contexts and the bark's position in the learned space, providing a strong proof-of-concept for the model's generalisation.

Despite these promising results, this study has limitations. We employed a foundational CNN with standard spectrograms, leaving the performance of more advanced architectures unexplored. The training was constrained by the Emotional Canines dataset, whose small scale and low benchmarks suggest inherent data ambiguity, limiting the performance ceiling of any model. Furthermore, our real-world validation was qualitative rather than a comprehensive, numerically-scored evaluation.

These limitations open avenues for future work. The learned space could be fine-tuned using larger, context-oriented datasets to anchor the coordinates to observable behaviours. Additionally, a detailed ablation study could disentangle the contributions of the base network architecture from the Siamese-style metric learning objective itself, leading to a more optimised framework for sparse, ordinally-annotated bioacoustics data which has large variability across dogs.

Beyond methodological refinements, practical directions include deployment in wearable monitoring systems, user-centered evaluation with handlers, and the incorporation of welfare-focused data collection protocols. Such translational steps are essential to bridge proof-of-concept modelling with the ethical, real-world demands of seizure-assistance applications.

## 6   Conclusions

This work successfully reframes canine vocalisation analysis from classification to a continuous regression task, introducing an adjusted Siamese Network that learns a semantically coherent arousal-valence space from sparse, ordinal data which varies across dogs. Our model outperforms baselines in preserving ordinal relationships, a critical requirement for meaningful dimensional embedding. To quantitatively measure this, we proposed the Turn-around Percentage (TAP) metric and demonstrated that our model achieves a significantly lower TAP, proving its superior ability to learn the correct ordinal structure.

A qualitative validation of our approach on a real-world dataset confirms the framework's generalisability. We provide a proof-of-concept for a methodology that moves beyond simple classification to enable a more nuanced, psychologically-grounded understanding of animal affect, laying the foundation for robustly detecting natural signaling behaviors in data-limited applications like seizure assistance. Future work should translate this framework into practical, welfare-aligned monitoring tools, extending its impact beyond proof-of-concept toward real-world assistance in detecting and alerting for seizure onset in the owners of trained assistance dogs.